\documentclass[12pt,titlepage]{article}
\usepackage{amssymb}
\usepackage{epsfig}
\usepackage{bm}
\usepackage{psfrag}

 \font\grande=cmr10 scaled
\outer\def\beginsection#1\par{\medbreak\bigskip
      \message{#1}\leftline{\bf#1}\nobreak\medskip
\vskip-\parskip
      \noindent}

\newcommand{\eq}{\begin{equation}}

\newcommand{\eqx}{\end{equation}}
\newcommand{\eqn}{\begin{eqnarray}}

\newcommand{\bi}{\begin{itemize}}

\newcommand{\eqnx}{\end{eqnarray}}
\newcommand{\ei}{\end{itemize}}
\newcommand{\ad}{{a^{\dagger}}}
\newcommand{\fd}{{f^{\dagger}}}
\newcommand{\bd}{{b^{\dagger}}}

\newcommand{\Qd}{{Q^{\dagger}}}

\newcommand{\nn}{\nonumber}
\newcommand{\ra}{\rangle}

\begin{document}
\titlepage

\begin{flushright}
\vspace{5mm}
 TPJU-13/2006
\end{flushright}
\vspace{10mm}
\begin{center}

\grande{ Solving some gauge systems at infinite N}

\vspace{15mm}

   \large{J. Wosiek}

   \vspace{2mm}

   {\sl M. Smoluchowski Institute of Physics, Jagellonian University}

{\sl Reymonta 4, 30-059 Krak\'{o}w, Poland}

\vspace{8mm}

\end{center}

\centerline{\medio  Abstract} \vskip 5mm \noindent

After summarizing briefly some numerical results for
four-dimensional supersymmetric SU(2) Yang-Mills quantum mechanics,
we review a recent study of systems with an infinite number of
colours. We study in detail a particular supersymmetric matrix model
which exhibits a phase transition, strong-weak duality, and a rich
structure of supersymmetric vacua. In the planar and strong coupling
limits, this field theoretical system is equivalent to a
one-dimensional XXZ Heisenberg chain and, at the same time, to a gas
of $q$-bosons. This not only reveals  a hidden supersymmetry in
these well-studied models; it also maps the intricate pattern of our
supersymmetic vacua into that of the now-popular ground states of
the XXZ chain.

  \vspace{5mm}

\begin{flushleft}
TPJU-13/2006\\
October 2006\\
\end{flushleft}

\newpage


\section{Introduction}

This lecture reviews a recent progress in studying the large N limit
of simple supersymmetric quantum mechanical systems which result
from the dimensional reduction of field theories with gauge
symmetry. Models of this type have been studied for a long time
\cite{CH,dWLN,JW1} and have many different applications, depending
on the space-time dimension, D, of the unreduced theory
\cite{CW1,CW2} (cf. Table \ref{roadmap}). For D=2 they
 can be often solved analytically \cite{CH,CW1,S,T}, providing quantitative
 realization of supersymmetry. In four space-time dimensions one of them
 is nothing but the small
volume limit of the Yang-Mills gluodynamics revealing the spectrum
of zero volume glueballas as a special case \cite{L,LM,vB,K}.
Finally, for D=10 they make contact with the M-theory via the BFSS
hypothesis \cite{BFSS}.

\begin{table}[h]
\begin{center}
\begin{tabular}{ccccc}
 \hline\hline
  $D$        &  $ 2 $ & $ 4 $ & $ \dots $ & $ 10 $  \\  \hline
  $N$        &        &       &           &         \\
   \hline
  2          &   $\bullet $    &    $\bullet $       &          &   $\circ$       \\
  3          &   $\bullet $       &        &          &          \\
  .          &   $\circ$     &        &          &          \\
  .          &   $\circ$      &        &          &          \\
  .          &        &        &          &          \\
  $\infty$   &   $\checkmark$     &        &          &  M     \\
   \hline\hline
\end{tabular}
\end{center}
\caption{A "road map" of a parameter space of supersymmetric
Yang-Mills quantum mechanics. Black dots denote models well
understood/solved, while open circles mark ones under study. A
checkmark labels parameters of a system discussed in this lecture.}
\label{roadmap}
\end{table}

Although we shall be mainly concerned with the large N limit, we
would like to give one example of the N=2 model.

\subsection{Four-dimensional supersymmetric Yang-Mills quantum mechanics at
finite N }
The system has three bosons and two fermions, both in the adjoint representation
of SU(2). The Hamiltonian reads \cite{CH}
\eqn H &=&  {1\over 2} p_a^i p_a^i + {g^2\over
4}\epsilon_{abc} \epsilon_{ade}x_b^i x_c^j x_d^i x_e^j + {i g
\over 2} \epsilon_{abc}\psi_a^{\dagger}\Gamma^k\psi_b x_c^k,
\label{eq:Hamiltonian} \nn \\
i=1,2,3&& a=1,2,3. \nn
\eqnx
The spectrum is obtained numerically by diagonalizing  $H$ in the gauge
 invariant eigen-basis of the occupation numbers of all degrees of freedom
 $n_a^i=\ad_a^i a_a^i, \eta_a^i=\fd_a^i f_a^i$
 \cite{JW1}
\eqn
|\{n_a^i,\eta_c^j\}>=\sum_{contractions}\ad^i_c {\ad}^j_d \ad^k_e \fd_b^m \fd_a^n \dots |0>. \nn
\eqnx
The Hilbert space was cut by restricting the gauge invariant total number
 of bosonic quanta
\eqn
B=\sum_{b,i} \ad_b^i a_b^i < B_{max}. \nn
\eqnx

\begin{figure}[ht]
\centering
 \psfig{figure=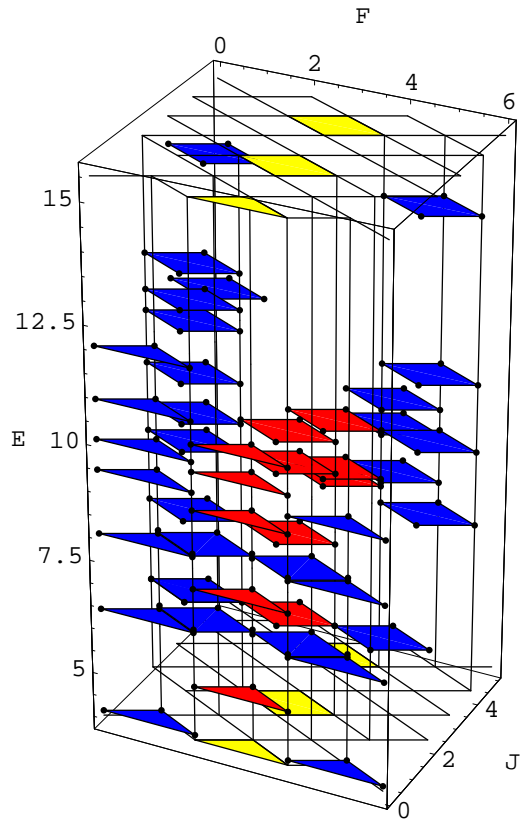,height=11cm}
 \caption{ The spectrum, and its supersymmetry structure, of the
D=4 supersymmetric Yang-Mills quantum mechanics.
\label{fig:3D}}
\end{figure}
It turned out that the physical (i.e. convergent with the cutoff
eigenenergies) could be computed for sizes of bases well within a
reach of a reasonably fast PC.
   The spectrum obtained in this way is shown in Fig.~\ref{fig:3D} \cite{CW2}.
It reveals a series of dynamical supermultiplets with degenerate
eigenenergies. The eigenstates within a supermultiplet are indeed
the supersymmetric images of one another. These states are localized
and they form a discrete spectrum. In addition there is also a
continuum. It occupies  central sectors of the figure, i.e. ones
with the conserved fermion number $F=2,3,4$. Supermultiplets of
non-localized states occur at every second value of the angular
momentum. Moreover, in these sectors the discrete and continuous
spectra coexists at the same energies. This unusual feature of gauge
interactions has never been observed so directly before. There are
two supersymmetric vacua with fermion numbers $F=2$ and $4$. This is
nothing but the zero-volume manifestation of the existence of the
$\lambda\lambda$ condensate in the space extended theories
\cite{JW2}. The condensate assumes two different (in fact opposite
for SU(2)) values, again in agreement with the non-trivial
predictions of the unreduced theory \cite{VY,NSVZ} .

\section{The large N limit and the planar calculus}
Above approach, even though quite successful, is naturally limited to the
low number of colours. For higher N the Hilbert spaces become too large.
Fortunately {\em at infinite N} Fock bases simplify enormously allowing
again to reach quantitative results \cite{VW1}-\cite{VW3}. This simplification is usually phrased
in terms of the Feynman diagrams and topological expansion \cite{tH}. However it can be
also formulated in the way suitable for the Hamiltonian formalism \cite{VW1,Adriano}.
 To this end
introduce matrix creation and annihilation (c/a) operators.
\eqn
a_{ik}=\sqrt{2} a^a T^a_{ik},\;\;\;f_{ik}= \sqrt{2} f^a
T^a_{ik},\;\;\; i,k=1,...,N. \nn \eqnx
The physical basis can be
conveniently generated by acting with the gauge invariant building
blocks (bricks)
\eqn (\ad \ad),(\ad\ad\ad),(\ad\ad\ad\ad),\dots,
(\ad^{N}),\;\;\;(.)\equiv Tr[.] \nn \eqnx
and their products, on the
empty Fock state $|0\ra$. Similarly for the fermionic sectors one
employs bricks with mixed fermionic and bosonic operators. It turns
out that in the large N limit only the single trace operators are
relevant. All products of traces are either non-leading or do not
provide new information (see \cite{PMO} for recent
results on that point). This observation culminates in the set
of simple rules to calculate explicitly matrix elements of various
Hamiltonians. One basically applies the Wick theorem employing the
commutation rules \eqn [a_{ik},\ad_{jl}]=\delta_{il}\delta_{kj}\nn
\eqnx and identifying the leading contributions. Details of such
"planar calculus" have been presented in \cite{VW1,Adriano,VW2}.
Here we summarize only two examples

A normalized state with n gluons in the F=0 sector reads
\eqn
|n>= \frac{1}{{\cal N}_n}Tr[(\ad)^n]|0>.  \label{state}
\eqnx
The normalization factor
\eqn
{\cal N}_n^2&=&<0|Tr[a^n] Tr[(\ad)^n]|0> \nn \\
&=&<0|(12)(23)...(n1)[1'2'][2'3']\dots[n'1']|0>,  \label{norm}\nn \\
(12)\equiv a_{i_1i_2},&\;\;\;&[12]\equiv \ad_{i_1i_2}.\nn
\eqnx
receives the maximal contribution only when the adjacent creation and annihilation
operators are contracted. This gives
\eqn
{\cal N}_n^2=n N^{n}.\nn
\eqnx
Similarly a matrix element of a typical term in a generic Hamiltonian
\eqn
 H_{n+2,n}=g^2 <n+2|Tr[\ad\ad\ad a]|n>  \label{hmn}\nn
\eqnx
can be explicitly calculated
\eqn
H_{n+2,n} =  g^2 N \sqrt{n(n+2)},&& \nn
\eqnx
and shown to depend only on the 't Hooft coupling $\lambda=g^2 N$.

   The rest of this lecture will be devoted to a rather simple system
which turns out to be surprisingly rich. The model may be considered
as a distant cousin of the $D=1+1$ supersymmetric
Yang-Mills theory reduced to the one point in space.

\section{A simple supersymmetric system }
Consider  one fermion and one boson with the following
supersymmerty generators and the Hamiltonian \cite{VW1}.
\eqn
Q= \sqrt{2} Tr [f \ad(1+g\ad)],&&
 \Qd = \sqrt{2} Tr [\fd (1+g a) a],\nn\\
H=&\{Q,\Qd\}&=H_B+H_F.\label{ham}
\eqnx
Explicitly
\eqn
H_B&=&\ad a + g(\ad^2 a + \ad a^2) + g^2 \ad^2 a^2,\nn \\
H_F&=& \fd f + g ( \fd f (\ad+a) + \fd (\ad+a) f) \nonumber \\
& + & g^2 ( \fd a f \ad + \fd a \ad f + \fd f \ad a + \fd \ad f a).\nn
\eqnx
This Hamiltonian conserves the gauge invariant fermion number
$F=Tr[\fd f]$ and can be diagonalized in sectors with well defined $F$.
For any finite N calculation of matrix elements of H quickly becomes cumbersome.
At infinite N, however, the planar rules illustrated above give simple and
compact expressions.
\subsection{Hamiltonian matrix at large N }
The gauge invariant basis in the $F=0$ sector can be chosen as (\ref{state}) with
 $n=0,1,2,3, ...$. Only $H_B$ contributes in this case. The matrix elements
in the planar approximation read
\eqn
<0,n|H|0,n>&=&(1+\lambda(1-\delta_{n1}))n, \nn \\
<0,n+1|H|0,n>=&<0,n|H|0,n+1>=&\sqrt{\lambda}\sqrt{n(n+1)}. \label{hmf1}
\eqnx
In the sector with one fermion the basis is
\eqn
|n>= \frac{1}{{\cal N}_n}Tr[(\fd\ad)^n]|0>,\;\;\;n=0,1,2,\dots . \nn
\eqnx
Now, and for higher $F$,
both $H_B$ and $H_F$ contribute. The Hamiltonian matrix is again simple
\eqn
<1,n|H|1,n>&=&(1 + \lambda)(n+1) +  \lambda,\nn \\
<1,n+1|H|1,n>=&<1,n|H_2|1,n+1>=&\sqrt{\lambda}(2+n). \label{hmf2}
\eqnx
The system reveals many interesting features already in these two lower sectors
\cite{VW1},
therefore we postpone the discussion of higher F's and turn to the physics of
ensembles with at most one fermion.

\subsection{Numerical results}

\begin{figure}[t]
\begin{center}
\epsfig{width=14cm,file=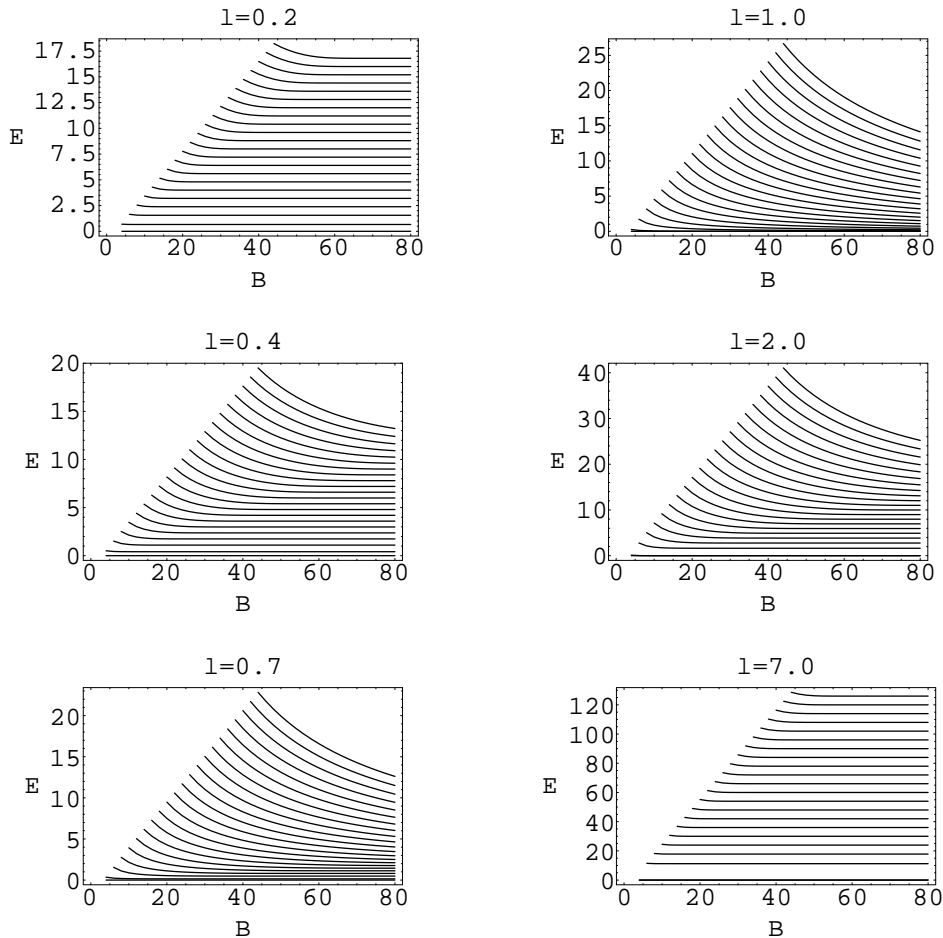}
\end{center}
\vskip-4mm
\caption{ The cutoff dependence of the spectra of $H$ in the F=0 sector, and
for a range of $\lambda$'s}
\label{fig:h2}
\end{figure}

\begin{figure}[tbp]
\psfrag{yyy}{${\scriptstyle E}$ }\psfrag{ttt}{${\scriptstyle\lambda=2.0}$}
\psfrag{xx1}{${\scriptstyle F=0\hspace*{2.5cm}F=1}$}\psfrag{tt1}{${\scriptstyle\lambda=0.5}$}
\psfrag{fff}{${\scriptstyle F=2\hspace*{2.5cm}F=3}$}
\epsfig{width=8cm,file=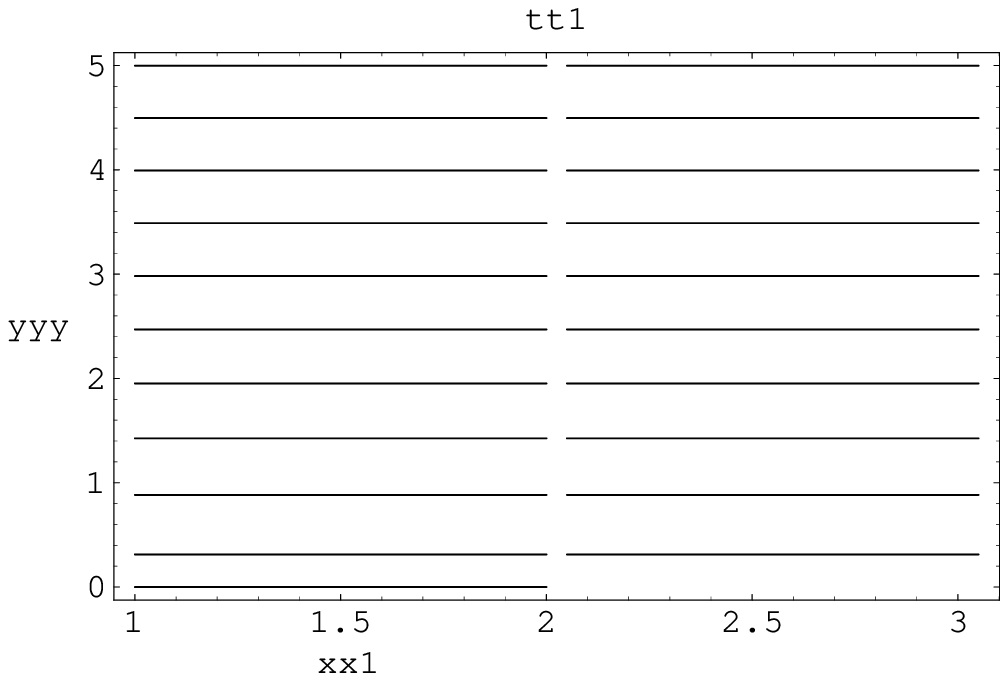}
\epsfig{width=8cm,file=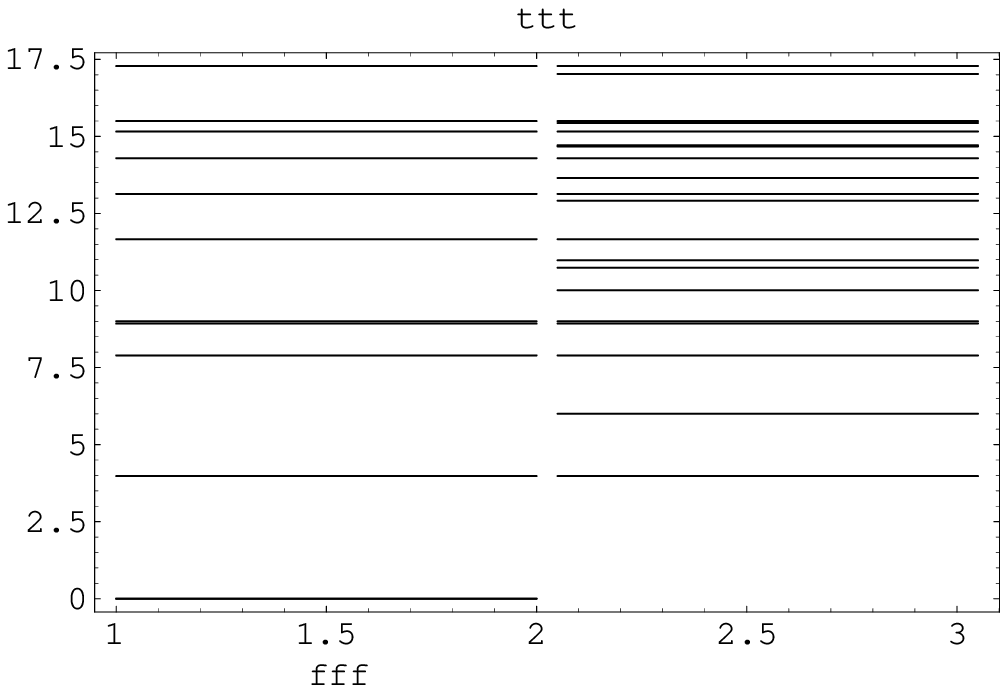}
\vskip-4mm \caption{Low lying bosonic and fermionic levels in the
first four fermionic sectors.} \label{fig:susyf01}
\end{figure}

The Hamiltonian matrix is sparse but infinite with rows and columns labeled by
the gauge invariant number of bosons $B=Tr[\ad a]$. To obtain,
the spectrum numerically we introduce the cutoff, $B_{max}$, limiting
the number of
bosonic quanta
 \eqn
  B < B_{max}, \nn
 \eqnx
and increase the cutoff until the spectrum converges. Results are shown in
Fig.\ref{fig:h2} for few values of the 't Hooft coupling. The convergence
with $B_{max}$ is satisfactory for $\lambda\ne 1$ and is faster for lower
eigenvalues. There, the limiting, "infinite volume", results can be easily
recovered. The less trivial way so see this is to check for the supersymmetry
which is broken by the cutoff and should be recovered only at infinite $B_{max}$.
Figure \ref{fig:susyf01} shows first few energy levels in four lower fermionic
sectors. The degeneracies between bosonic and fermionic partners of
the $F=(0,1)$ supermultiplets
are excellent. Moreover, there is also an unbalanced supersymmetric vacuum state
with zero eigenenergy \footnote{An apparent lack of the degeneracy between some
states with higher $F$ will be explained later.}. These results provide also the
non-trivial test of our planar rules: the planar approximation does not break
supersymmetry.

The slowing of the cutoff dependence around $\lambda=1$ is a characteristic
signature of a phase transition. At this point the system looses the energy gap
and the spectrum becomes continuous. Any finite number of low-lying levels
collapses to zero at infinite cutoff, but the cutoff dependence becomes
characteristic of that for the continuous spectrum.

\begin{figure}[t]
\begin{center}
\epsfig{width=8cm,file=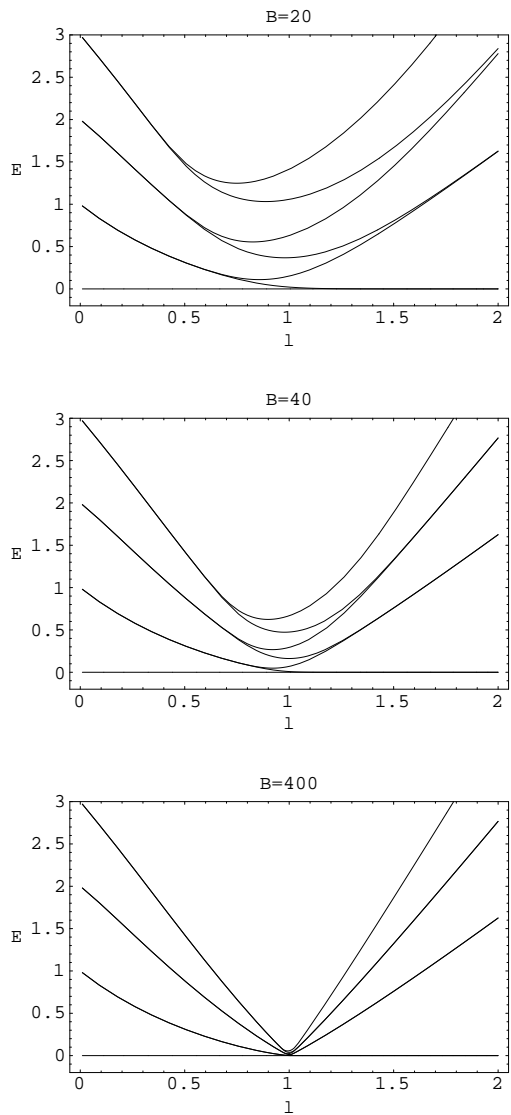}
\end{center}
\vskip-4mm
\caption{Lowest bosonic and fermionic levels, as functions of $\lambda$, for different cutoffs}
\label{fig:rearr}
\end{figure}

Another interesting feature of this phase transition is the rearrangement
of members of supermultiplets. This is shown in Fig.\ref{fig:rearr} where
the $\lambda$ dependence of the first four eigenenergies from both sectors
is displayed for few values of $B_{max}$.
Away from the critical region supersymmetry is quickly restored and the F=0
and F=1 levels are undistinguishable even for low cutoffs. Close to the criticality
 partners do not have the same energies since SUSY is broken at finite
cutoff. Interestingly however, they {\em rearrange} while the 't Hooft coupling
passes its critical value. In particular, the {\em new} vacuum state appears
when $\lambda$ moves from the low to the large coupling phases. This also
implies (and was readily found) that the Witten index
(restricted to the $F=0,1$ sectors) jumps by one unit across the
phase transition point.

Finally, the system has the exact strong-weak duality
\eqn
b \left(E^{(F=0)}_n(1/b) -\frac{1}{b^2}\right)  &=&
\frac{1}{b} \left( E^{(F=0)}_{n+1}(b) -b^2 \right), \nn \\
b \left(E^{(F=1)}_n(1/b) -\frac{1}{b^2}\right)   &=&
 \frac{1}{b} \left( E^{(F=1)}_n(b)   -b^2 \right) . \nn
\eqnx
It follows directly from the matrix representation (\ref{hmf2}) for $F=1$ .
However it is not obvious in the $F=0$ sector, cf. Eq. (\ref{hmf1}),
but it is a direct consequence of supersymmetry.

\subsection{Analytic solution}
All above results have been subsequently derived analytically \cite{VW1}. For example
the second vacuum state has the form $b \equiv \sqrt{\lambda}$
\eqn
|0\rangle_2 = \sum_{n=1}^{\infty} \left(\frac{-1}{b}\right)^{n} \frac{1}{\sqrt{n} }|0,n\rangle \ .
\label{vac2}
\eqnx
It is indeed annihilated by $\Qd$,
and exists only for $ b > 1 $.

Surprisingly, one can find analytically the complete spectrum and construct
all eigenstates in the $F=0$ sector. To this end it is convenient to
introduce another, not orthonormal basis,
\eqn
  |B_n\rangle &=&\sqrt{n}|n\rangle+b\sqrt{n+1}|n+1\rangle \, . \nn
\eqnx

This basis is convenient since the action of the Hamiltonian on $|B_n\rangle$
is so simple that the generating function for the expansion of the eigenstates
in that basis
\eqn
f(x)=\sum_{n=0}^{\infty} c_n x^n \, \;\;\;\;\;\;\leftrightarrow\;\;\;\;\;\;
|\psi\rangle = \sum_{n=0}^{\infty} c_n |B_n\rangle \nn
 \label{spsi}
\eqnx
can be constructed. The solution reads
\eqn
f(x)&=& \frac{1}{\alpha}\;\frac{1}{x+1/b}\;  F(1,\alpha;1+\alpha;\frac{x+b}{x+1/b}),\;\;\; b < 1, \nn\\
f(x)&=& \frac{1}{1-\alpha}\;\frac{1}{x+b}\; F(1,1-\alpha;2-\alpha;\frac{x+1/b}{x+b}),\;\;\; b > 1, \nn\\
E &=& \alpha(b^2-1), \nn
\eqnx
and together with the quantization condition
\eqn
f(0)=0,
\eqnx
reproduces numerical eigenvalues obtained earlier in the infinite cutoff limit.

 As a one check, set $\alpha=0$ in the generating function, for $b > 1$, to obtain
\eqn
f_0(x)=\frac{1}{1+b x} \log\frac{b+x}{b-1/b}, \;\;\;  b>1,
\eqnx
which indeed reproduces the second ground state (\ref{vac2}).

\section{Higher fermionic sectors: F=2,3}

Single trace states with two fermions
\eqn
|n_1,n_2\ra & = & \frac{1}{{\cal N}_{n_1 n_2}}
Tr[\ad^{n_1}\fd\ad^{n_2}\fd ]|0\ra,\nn
\eqnx
are labeled by two integers whose ordering
is important modulo a cyclic permutation. Pauli principle eliminates
states with $n_1=n_2$, hence we can always take $n_2 < n_1$. The planar rules
give for the Hamiltonian matrix in this sector \cite{VW2}
\eqn
\langle n_1,n_2|H|n_1,n_2\rangle& = &(n_1+n_2+2)(1+b^2) -
b^2 (2 - \delta_{n_1,0}) - 2 b^2 \delta_{n_2,n_1+1},\nn \\
\langle n_1+1,n_2|H|n_1,n_2\rangle = &b (n_1+2)& = \langle n_1,n_2|H|n_1+1,n_2\rangle,\nn\\
\langle n_1,n_2+1|H|n_1,n_2\rangle = &b (n_2+2)& = \langle n_1,n_2|H|n_1,n_2+1\rangle.\nn\\
\langle n_1+1,n_2-1|H|n_1,n_2\rangle  &=& \langle n_1,n_2|H|n_1+1,n_2-1\rangle
\nn\\&=& 2 b^2 (1- \delta_{n_2,n_1+1}).\nn
\eqnx
 Planar  states in the three fermion sector are labeled by three integers
modulo cyclic translations
\eqn
|n_1,n_2,n_3\ra & = & \frac{1}{{\cal N}_{n_1 n_2 n_3}}Tr[\ad^{n_1}\fd\ad^{n_2}\fd \ad^{n_3}\fd ]|0\ra,\nn\\
 0\leq n_1 ,\;\;\;\;&& n_1 \leq n_2,\;\;\;\;\ n_1 \leq n_3.  \nn
\eqnx
Again the Hamiltonian matrix was explicitly calculated
\eqn
\langle n_1,n_2, n_3|H|n_1,n_2,n_3\rangle = (n_1+n_2+n_3 +3)(1+b^2) \nn \\
- b^2 (3 - \delta_{n_1,0} -
\delta_{n_2,0}- \delta_{n_3,0})\nn
\label{diag3} ,
\eqnx
\eqn
\langle n_1+1,n_2,n_3|H|n_1,n_2,n_3\rangle &=b (n_1+2) \Delta=& \langle n_1,n_2,n_3|H|n_1+1,n_2,n_3\rangle
,   \nn\\
{\rm plus ~ cyclic} \nn
\label{off31}
\eqnx
\eqn
\langle n_1+1,n_2-1,n_3|H|n_1,n_2, n_3\rangle  &= b^2 \Delta=& \langle n_1,n_2,n_3|H|n_1+1,n_2-1, n_3\rangle ,
\nn\\{\rm plus ~ cyclic} \nn
\label{off32}.
\eqnx
where $\Delta=1/\sqrt{3}$ if $n_1=n_2=n_3$, and $\Delta=\sqrt{3}$ if the final state is of this form,
otherwise $\Delta=1$.

    Numerical computation of the spectrum proceeds as before.
There is again a phase transition causing the critical
slowing down around $\lambda_c=1$. Away from it, the eigenenergies converge
satisfactorily, with the infinite volume results shown in Fig.\ref{fig:susyf01}.
   The spectrum in higher fermionic sectors is essentially different from
the $F=0,1$ case. We again find dynamical $F=(2,3)$ supermultiplets, however
not all states with three fermions have their superpartners in the $F=2$ sector.
Instead, they are degenerate with ones from the $F=4$ sector. This pattern continues
now for all F as follows from counting the number of states: it is an increasing
function of F. Another novel feature is that the spectrum is rather irregular,
while the eigenenergies are almost equidistant for $F=0,1$.
    There is again the rearrangement of members of supermultiplets across
the phase transition. Interestingly however, the {\em two} new SUSY vacua
with $F=2$ appear
in the strong coupling phase, while there is none at weak coupling.

    Similarly to the $F=0$ case, the two vacua can be constructed analytically.
To this end consider the "extreme" strong coupling limit of the Hamiltonian
(\ref{ham})\cite{VW2}
\eqn
 H_{SC}=\lim_{\lambda\rightarrow\infty} \frac{1}{\lambda} H  =  Tr (\fd f) +
 \frac{1}{N}[Tr(\ad^2 a^2) + Tr(\ad \fd a f) + Tr(\fd\ad f a)].
 \label{hstr}
\eqnx
Surprisingly, $H_{SC}$ conserves the number of bosonic quanta as well, and proves
very useful in mapping the structure of the model in {\em all} fermionic sectors
(cf. the following Section). Coming back to $F=2$, one can identify the
finite dimensional
$(F=2,B)$ sectors with zero eigenvalues, and construct the corresponding eigenstates.
They read
\eqn
|F=2,v\ra_1^{\infty} &=& |0,1\ra, \\
|F=2,v\ra_2^{\infty} &=& |0,3\ra - 2|1,2\ra,
\eqnx
the superscript referring to the infinite value of the 't Hooft coupling.
These states can then be used to construct the two vacua at finite $\lambda > 1$
\eqn
|F=2,v\rangle_1^{\lambda} &=& \sum_{n=1}^{\infty} (-\sqrt{\lambda})^{-n}
|0,n\rangle ,\ \nn \\
|F=2,v\rangle_2^{\lambda} &=& (1+Q_s H_s^{-1} \Qd_w)^{-1}|F=2,v\ra_2^{\infty},
\eqnx
with $Q_s=g Tr[f \ad^2],\; H_s=\{Q_s,\Qd_s\},\; \Qd_w = Tr[\fd a]$ \cite{VW2}.

\begin{figure}[tbp]
\psfrag{yyy}{$q_n+n-1$}\psfrag{xxx}{$B_{max}$}
\begin{center}
\epsfig{width=8cm,file=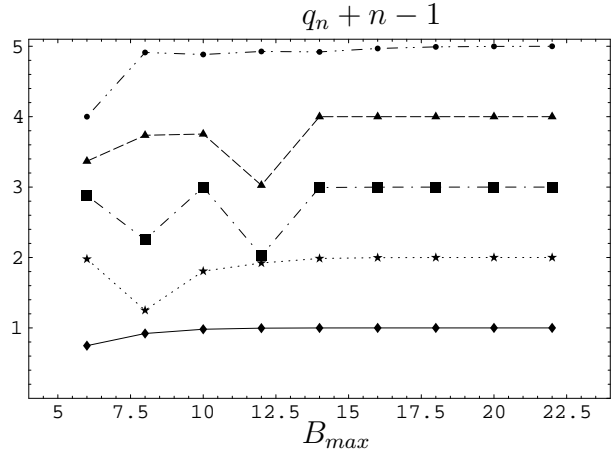}
\end{center}
\vskip-4mm \caption{First five supersymmetry fractions.} \label{fig:sfrac}
\end{figure}

The restoration of supersymmetry can also bee seen on the level of eigenstates.
Members of supermultiplets transform among themselves by SUSY charges
$Q$ and $\Qd$, which become the well defined matrices in the planar bases. To see
SUSY in this way define the following supersymmetry fractions
\eqn
q_{m n}\equiv \sqrt{\frac{2}{E_m+E_n}} < F+1,E_m|Q^{\dagger}|F,E_n > ,
\eqnx
which are the coordinates of the supersymmetric images of eigenstates
in the corresponding fermionic sectors. Fig.\ref{fig:sfrac} shows that
SUSY fractions indeed quickly stabilize with incrasing $B_{max}$.

\begin{figure}[tbp]
\psfrag{yyy}{${\scriptstyle I_W(6)}$}\psfrag{xxx}{${\scriptstyle \sqrt{\lambda}}$}
\begin{center}
\epsfig{width=8cm,file=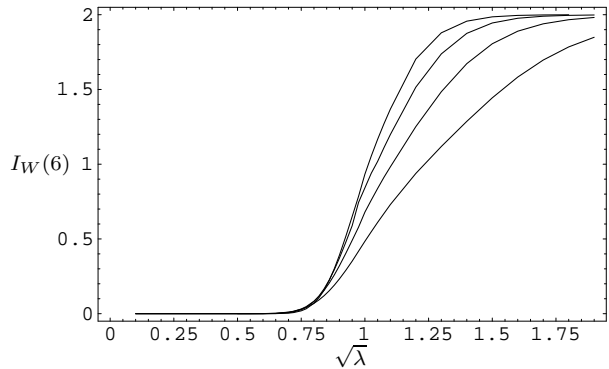}
\end{center}
\vskip-4mm \caption{Behaviour of the restricted Witten index,
at $T=6$, around the phase transition.} \label{fig:wi}
\end{figure}

Supersymmtery fractions are also useful in defining the restricted Witten index
which smoothly interpolates, at finite cutoff, between the two phases. The straightforward
restriction of the sum
\eqn
W(T,\lambda)= \sum_i (-1)^{F_i} e^{-T E_i} , \nn
\eqnx
to the $F=2,3$ eigenstates does not suffice since some of the states with three
fermions remain unbalanced. Instead, one can define
\eqn
W_R(T,\lambda)= \sum_i  \left( e^{-T E_i} -  e^{-T \bar{E}_i} \right),\;\;\;
\bar{E}_i=\frac{\sum_f E_f |q_{fi}|^2}{\sum_f |q_{fi}|^2},  \nn
\eqnx
i.e. we take as the energy of the supersymmetric partner the energy weighted
by supersymmetric fractions. This definition enforces summation only over
complete supermultiplets away from the critical region, while provides
the smooth smearing among possible candidates around $\lambda_c$.
The index defined this way changes smoothly at finite cutoff, and varies
 by two units
as expected (cf. Fig.\ref{fig:wi}) \cite{VW2}.

\section{Arbitrary F}

A simple generalization of the $F=2,3$ cases for arbitrary number of fermions
reads \cite{OVW1}
\eqn
|n\rangle=|n_1,n_2,\dots,n_F\rangle =
\frac{1}{{\cal N}_{\{n\}}}
Tr(\ad^{n_1}\fd \ad^{n_2}\fd\dots\ad^{n_F}\fd)|0\ra   \label{anyF}
\eqnx
So the states of the basis may be labeled by $F$ bosonic occupation numbers
$\{n_1,n_2,\dots n_F\}$ (configurations) modulo cyclic shifts. Some states
are excluded by the Pauli principle. For example
\eqn
\{3,3\},\;\;\;\;\;{\rm or}\;\;\;\;\;\{2,1,1,1,2,1,1,1\}  \nn
\eqnx
are not allowed, since they change sign while retaining their identity after
the suitable number of cyclic shifts. When calculating normalizations and
matrix elements one should keep track of symmetry factors
defined as the number of cyclic shifts which bring a state back to itself.
For example the state $|2,1,2,1,2,1\ra$ has the symmetry factor $d=3$.
Alternatively one can label planar states with the periodic binary strings
(necklaces) with, e.g., zeros/ones corresponding to the bosonic/fermionic
creation operators (for a more complete discussion see \cite{OVW1}).

    Detailed calculation of spectra for arbitrary fermion number
is now in progress. However many properties, e.g. the supersymmetry structure,
has been inferred from the strong coupling limit of the model.

As already mentioned, the strong coupling Hamltonian (\ref{hstr}) conserves
both fermionic and bosonic numbers of quanta. Hence the Hilbert space
splits into the finite dimensional sectors, cf. Table \ref{bastab},
and the Hamiltonian in each sector becomes a finite matrix. Still the
$H_{SC}$ has the exact supersymmetry with the strong-coupling charges
\eqn
Q_{SC}=\frac{1}{\sqrt{N}} Tr[f\ad^2],\;\;\;\Qd_{SC}=\frac{1}{\sqrt{N}} Tr[f\ad^2],
\eqnx
acting along the "diagonals" $2F+B=const.$ of Table \ref{bastab}.
Diagonalizing few of these matrices we have indeed found expected
degeneracies\footnote{Obviously no cutoff was required this time.}. Interestingly,
we have also found experimentally \cite{OVW1,VW3} that the zero energy eigenstates
are located {\em only} in the sectors with
\eqn
B=F \pm 1,\;\;\;F\; even.  \label{magic}
\eqnx
which form the regular "magic" staircase, shown in the boldface, in Table \ref{bastab}.

\begin{table}[tbp]
\begin{center}
{
\begin{tabular}{||c|ccccccccccc}
$11\;\;\;$ & 1  &  1 & 6 &  26  & 91 &  273 & 728 & 1768 & 3978 & 8398  & {\bf 16796} \\
$10\;\;\;$ & 1  &  1 & 5 &  22  & 73 &  201 & 497 & 1144 & 2438 & 4862  & 9226 \\
$9\;\;\;$ & 1  &  1 & 5 &  19  & 55 &  143 & 335 & 715 & {\bf 1430} & 2704  & {\bf 4862} \\
$8\;\;\;$ & 1  &  1 & 4 &  15  & 42 &  99 & 212 & 429 & 809 & 1430  & 2424 \\
$7\;\;\;$ & 1  &  1 & 4 &  12  & 30 &  66 & {\bf 132} & 247 & {\bf 429} & 715  & 1144 \\
$6\;\;\;$ & 1  &  1 & 3 &  10  & 22 &  42 & 76 & 132 & 217 & 335  & 497 \\
$5\;\;\;$ & 1  &  1 & 3 &  7  & {\bf 14} &  26 & {\bf 42} & 66 & 99 & 143  & 201 \\
$4\;\;\;$ & 1  &  1 & 2 &  5  & 9 &  14 & 20 & 30 & 43 & 55  & 70 \\
$3\;\;\;$ & 1  &  1 & {\bf 2} &  4  & {\bf 5} &  7 & 10 & 12 & 15 & 19  & 22 \\
$2\;\;\;$ & 1  &  1 & 1 &  2  & 3 &  3 & 3 & 4 & 5 & 5  & 5 \\
$1\;\;\;$ & {\bf 1}  &  1 & {\bf 1} &  1  & 1 &  1 & 1 & 1 & 1 & 1  & 1 \\
$0\;\;\;$ & 1  &  1 & 0 &  1  & 0 &  1 & 0 & 1 & 0 & 1  & 0 \\
\hline 
$B\;\;\;$        &  $  $ & $  $ & $  $
  & $  $ &  $  $ & $  $ & $  $  & $  $ &  $  $ & $  $ & $  $ \\

  $\;\;\;F$        &  $ 0 $ & $ 1 $ & $ 2 $
  & $ 3 $ &  $ 4 $ & $ 5 $ & $ 6 $  & $ 7 $ &  $ 8 $ & $ 9 $ & $ 10 $ \\

   \hline\hline
\end{tabular}
}
\end{center}
\caption{
Sizes of gauge invariant bases in the $(F,B)$ sectors. }
\label{bastab}
\end{table}

This observation, when combined with the weak coupling (harmonic oscillator)
limit, explains the structure of SUSY vacua for all $F$
and at any value of the 't Hooft coupling. Namely, for any even $F$ there
are two supersymmetric vacua for any $\lambda$ in the strong coupling phase.
On the other hand, there is only one SUSY vacuum in the weak coupling
 phase and it has $F=0$.

The question {\em why} the supersymmetric vacua are located
in the magic sectors (\ref{magic}) will be answered in the next Section.

\section{Two equivalencies with statistical systems}
Interestingly, our strong-coupling model is equivalent to the two
well known and nontrivial statistical systems \cite{VW3}.
\subsection{A gas of q-bosons}
Consider a one dimensional, periodic lattice with size $F$.
At each lattice site put a bosonic degree of freedom described by its
c/a operators $\ad_i,a_i$.
The strong coupling Hamiltonian (\ref{hstr}) is equivalent to the following
one expressed solely in terms of bosonic variables
\eqn
H=B + \sum_{i=1}^F \delta_{N_i,0} + \sum_{i=1}^F b_i \bd_{i+1} + b_i \bd_{i-1},
\label{hlat}
\eqnx
where $N_i=\ad_i a_i$,  $B=N_1+N_2+\dots +N_F$, and the action of $b_i, \bd_i$
is
\eqn
\bd|n\ra = |n+1\ra,\;\;\; & b|n\ra = |n-1\ra,&\;\;\; b|0\ra\equiv 0, \label{qbos}
\eqnx
that is, they create and annihilate one quantum {\em without} the usual
$\sqrt{n}$ factors. They satisfy the following commutation rules
\eqn
 [b,\bd]=\delta_{N,0}
\eqnx
with the same non-linear (in b's) $\delta$ operator as in the Hamiltonian.
Using the planar rules discussed earlier, one can show that the action
of the first two terms of (\ref{hstr}) on the planar basis is exactly the same
as the action of the first two terms of (\ref{hlat}) in the eigenbasis of $N_i$ \cite{VW3}.
The last two terms of (\ref{hstr}) are the same as the hopping terms of
(\ref{hlat}).  Since the planar states acquire a phase $(-1)^{F-1}$ upon
 cyclic shifts, the planar system has the same spectrum as (\ref{hlat})
 in the eigen-sector of the lattice shifts, $U$, with $\lambda_U=(-1)^{F-1}$.
This equivalence was cross-checked numerically for $3\le F, B \le 7$.

The c/a operators are known in the literature \cite{QB1,QB2}. Their algebra is the special case
of the q-deformed harmonic oscillator algebra
\eqn
 [b,\bd]=q^{-2N}
\eqnx
with $q\rightarrow\infty$. Transitions (\ref{qbos}) without the square roots
are often referred to as {\em assisted} (i.e. independent of the occupancy) transitions.
The system of q-bosons described by the Hamiltonian (\ref{hlat}) is,
 to our knowledge, regarded as non-soluble. In view of the second equivalence,
 discussed below, it turns out to be in fact soluble.

\subsection{The XXZ Heisegberg chain}
The second equivalence has been proved employing another representation
of the planar states (\ref{anyF}) \cite{VW3}
\eqn
 \frac{1}{{\cal N}_n}|
Tr[\ad^{m_1}(\fd)^{n_1}\ad^{m_2}(\fd)^{n_2} \dots (\fd)^{n_k}]|0\ra \equiv
\left| (0)^{m_1}(1)^{n_1}\dots (0)^{m_r}(1)^{n_r} \right\ra  \,
r\ge 1,\;m_i, n_i >0) \, ,
\label{plstates}
\eqnx
Where the states are labeled by the binary strings with, e.g. 0/1 corresponding
to
the bosonic/fermionic creation operators. In this basis, the action of
the strong coupling Hamiltonian (\ref{hstr}) is equivalent to that of the XXZ
chain
\eqn
H_{\rm XXZ}^{(\Delta)} &=& -\frac12 \sum_{i=1}^L \left( \sigma_i^x \sigma_{i+1}^x
+ \sigma_i^y \sigma_{i+1}^y + \Delta~    \sigma_i^z \sigma_{i+1}^z    \right) \nonumber
\eqnx
for the particular values of the anisotropy parameter $\Delta$.
The detailed correspondence reads
\eqn
H_{{\rm SC}}(F,B)=\left\{\begin{array}{ccc}
             -H_{\rm XXZ}^{(+ 1/2)}+\frac{3}{4}L \ ,\   & F ~{\rm  odd}, &\\
             +H_{\rm XXZ}^{(- 1/2)}+\frac{3}{4}L \ , \  & F ~{\rm  even}, B~ {\rm odd} \, ,
            \end{array}
            \right.              \label{xxzeq}
\eqnx
where the spin Hamiltonians are restricted to the translationally invariant
sectors with the lattice size and the total spin fixed by $(F,B)$:
\eqn
L=F+B,\;\;\;\;S^z=\sum_{i=1}^L s_i^z=\frac{1}{2}(F-B),\;\;\;\label{xxzpar}
\eqnx
Remarkably, this equivalence implies the existence
of the magic staircase of the supersymmetric vacua discussed above.
More than thirty years ago Baxter has found that, for $\Delta=-1/2$,
the ground states with $S_z=\pm 1/2$ have particularly simple eigenenergy
$E_0=-\frac{3}{4}L$ for infinite L \cite{Ba1}. Recently his findings
have been extended by Riazumov and Stroganov to any finite, odd L \cite{RS}.
 In view of (\ref{xxzeq},\ref{xxzpar})
the Riazumov-Stroganov states are nothing but our strong coupling vacua \cite{VW3}.
 Moreover, supersymmetry
requires a host of degeneracies among the, seemingly unrelated,
excitations of the XXZ chain. Amusingly, since the supersymmetry
transformations change the lattice size, these excitations live on
different lattices! \footnote{Supersymmetry of the, $\Delta=-1/2$,
XXZ chain and other statistical systems, has been discussed in the
literature, albeit in slightly different contexts (see, e.g.
\cite{FNS} and references therein). We thank Jan de Gier for
bringing this to our attention. }

Finally, since the XXZ model is exactly soluble (e.g. by the Bethe Ansatz \cite{Fad}),
our supersymmetric system is also soluble at infinite coupling.
As one application we give here the algebraic determination of the Bethe
phase factors for the first three steps of the magic staircase.
The eigenenergies of $H_{XXZ}(\Delta)$ are \cite{XXZ}
\eqn
E_{XXZ}(\Delta)=-L\frac{\Delta}{2} + 2 m \Delta -2 \sum_{j=1}^m \cos{p_j}, \label{baen}
\eqnx
where the momenta $ -\pi < p_j < \pi $ satisfy the following set of Bethe equations
\eqn
e^{i L p_j} = (-1)^{m-1} \prod_{l=1}^m e^{i(p_j-p_l)}
\frac{e^{i p_l}+e^{-i p_j} - 2\Delta}{e^{i p_j}+e^{-i p_l} - 2\Delta},\;\;\;
j=1,\dots,m. \label{beqs}
\eqnx
With $m$ denoting the number of down spins in a chain.
For the supersymmetric model $m=B$ and (\ref{baen}) translates into
\eqn
E_{SC}(F,B)=&F+2\sum_{j=1}^B \cos{p_j},& \;\;\;{\rm for}\; F\;\;{\rm odd}~,~
\Delta = + \frac12 ,\\
E_{SC}(F,B)=&F-2\sum_{j=1}^B \cos{p_j},& \;\;\;{\rm for}\; F\;\;{\rm even},
\;\;{\rm and}\;\;\ B\;\; {\rm odd}
~,~ \Delta = -  \frac12 .
\label{SBAeven}
\eqnx
Solving numerically Eqs.(\ref{beqs}) we have found that the supersymmetric
vacua occur at F even and B odd, and are always given by the yet simpler sub-ansatz of the Bethe Ansatz
\eqn
p_1=0,\;\;\; p_{2k+1}=-p_{2k}, \;\;\; k=1,...,(B-1)/2.    \label{an2}
\eqnx
which reduces the number of independent variables roughly by a factor of 2.
Still, the Bethe equations can only be solved numerically in general. However,
for the first three sectors the problem can be managed algebraically.
We shall discuss it separately for each sector.
\subsubsection{F=4, B=3}
This corresponds to $L=7$, $m=3$, and Eq.(\ref{an2}) implies
\eqn
p_1=0,\;\; p_2=p,\;\;p_3=-p,
\eqnx
 and (\ref{beqs}) reduces to ($z=e^{ip}$)
\eqn
z^6=1.
\eqnx
The admissible solution, which gives $E_{SC}=0$, is
\eqn
z=e^{i\frac{\pi}{3}}.
\eqnx
\subsubsection{F=4, B=5}
Now $L=9$, $m=5$, and with the aid of (\ref{an2}) Bethe equations reduce to
($x=e^{i p_2},\;\;y=e^{i p_4}$)
\eqn
x^8 & = & \frac{x y + x + y}{x+y+1}\;\frac{x y + x + 1}{x y + y + 1}, \nn \\
y^8 & = & \frac{x y + x + y}{x+y+1}\;\frac{x y + y + 1}{x y + x + 1}.  \label{beq45}
\eqnx
Solving (\ref{beq45}) is difficult in general, however one can easily find solutions with
vanishing energy (\ref{SBAeven}).
\eqn
E_{SC}(4,5)=2-2\left(x+y+\frac{1}{x}+\frac{1}{y}\right) = 0. \label{ben}
\eqnx
 Introducing two symmetric variables
\eqn
s=x+y,\;\;\;\;p=x y,
\eqnx
one obtains from (\ref{beq45}), (\ref{ben}) the following  two equations
for $s$ and $p$
\eqn
 \frac{s+p}{s+1}&=& \pm p^4, \\
s+\frac{s}{p}&=&1.
\eqnx
The admissible solution is on the negative branch and reads
\eqn
x=\frac{1}{64}\left(16+i \sqrt{2}\sqrt{15+\sqrt{33}}(7-\sqrt{33})
-4\sqrt{-16(3+\sqrt{33})-i 2\sqrt{2}\sqrt{15+\sqrt{33}}(9+\sqrt{33})}\right),\nn \\
y=\frac{1}{64}\left(16+i \sqrt{2}\sqrt{15+\sqrt{33}}(7-\sqrt{33})
+4\sqrt{-16(3+\sqrt{33})-i 2\sqrt{2}\sqrt{15+\sqrt{33}}(9+\sqrt{33})}\right). \nn \\  \label{xy45}
\eqnx
This pair indeed satisfies both Bethe equations, together with (\ref{ben}),
and therefore corresponds to our supersymmetric vacuum.
\subsubsection{F=6, B=5}
This sector corresponds to $L=11$, $m=5$.
Now the reduced Bethe equations are
\eqn
x^{10} & = & \frac{x y + x + y}{x+y+1}\;\frac{x y + x + 1}{x y + y + 1}, \nn \\
y^{10} & = & \frac{x y + x + y}{x+y+1}\;\frac{x y + y + 1}{x y + x + 1}.  \label{beq65}
\eqnx
As previously, we look for the solutions which satisfy $E_{SC}(6,5)=0$, that is
\eqn
x+y+\frac{1}{x}+\frac{1}{y} = 2,\;\;\;{\rm or}\;\;\;\frac{s}{p}=\frac{2}{1+p}.
\eqnx
The second equation follows from the product of the Bethe equations
\eqn
p^5=\pm \frac{p+s}{1+s},\;\;\;{\rm or}\;\;\; \frac{s}{p}=-\frac{1 \mp p^4}{1 \mp p^5}.
\eqnx
These equations reduce to the fourth order polynomial equation for p.
Again there is only one admissible solution and it lies on the negative branch
\eqn
x=\frac{1}{72}\left(36+i \sqrt{2}\sqrt{11+\sqrt{13}}(7+\sqrt{13})
-6\sqrt{2}\sqrt{6(-3+\sqrt{13})+i \sqrt{2}\sqrt{11+\sqrt{13}}(-5+\sqrt{13})}\right),\nn \\
y=\frac{1}{72}\left(36+i \sqrt{2}\sqrt{11+\sqrt{13}}(7+\sqrt{13})
+6\sqrt{2}\sqrt{6(-3+\sqrt{13})+i \sqrt{2}\sqrt{11+\sqrt{13}}(-5+\sqrt{13})}\right). \nn \\  \label{xy65}
\eqnx

As the last exercise let us show that these numbers are indeed unimodular which
is not obvious at first sight. However it is easy to calculate
\eqn
|x+y|^2=\frac{1}{3}(7+\sqrt{13}),\;\;|x-y|^2=\frac{1}{3}(5-\sqrt{13}),\;\;
{\rm therefore} \;\;|x|^2+|y|^2=2.  \label{unimod}
\eqnx
Further, since $x=a+b$ and $y=a-b$, one can observe that
\eqn
(a b^*)^2=\frac{1}{72}(-11+\sqrt{13}),\;\;\Rightarrow\;\; |x|^2=|y|^2 ,
\eqnx
which together with (\ref{unimod}) implies that $x$ and $y$ are indeed
pure phase factors. As a byproduct one obtains
\eqn
Re(x y^*)\equiv \cos{(p_2-p_4)}=\frac{1}{6}(1+\sqrt{13}).
\eqnx
\section{Discussion}
The direct diagonalization of a Hamiltonian matrix is usually
considered a viable tool, for finding a spectrum, only for finite
matrices. It turns out however, that the approach works in many
cases with the infinite dimensional Hilbert space as well. Although
such Hilbert spaces appear already in many quantum mechanical
systems with finite number of degrees of freedom, the true challenge
for the Hamiltonian formalizm is posed by the field theoretical
problems. An interesting family of systems results from the
dimensional reduction of field theoretical models to a one
point in space. As such, they again have the finite number of degrees
of freedom, however they inherit many advanced features of
parent field theories, for example their
 symmetries including
supersymmetry. The straightforward diagonalization proved quite successful in uncovering
quite rich and nontrivial spectra of supersymmetric Yang-Mills quantum mechanics in various dimensions.

This lecture reviews the recent study of a system with an infinite
number of degrees of freedom, namely a particular supersymmetric gauge quantum mechanics
(also referred to as a matrix model) with the infinite number of colours.
The model was conceived as the illustration of the planar calculus in the Fock
space. However it was subsequently found that the system has an interesting
physics which connects to many recently discussed issues. For example, the system
undergoes the discontinuous phase transition in the 't Hooft coupling which is
accompanied
by the remarkable rearrangement of dynamical supermultiplets. It enjoys the strong-weak
duality in the lowest (and simplest) fermionic sectors where the complete,
exact spectrum can be found analytically. The full structure of intervening
supermultiplets begins with two fermions and goes on ad infinitum.
In each bosonic  sector two new supersymmetric vacua appear in the strong
coupling phase while there is only one in the weak coupling region.
The behaviour in the strong coupling phase has been found by studying the system
at the infinite value of the 't Hooft coupling and then extending it to the whole
strong coupling regime.

At infinite coupling the model reveals also its connections
with the statistical physics, which proves, among other things, that
a quantum mechanics at infinite $N$ becomes a {\em bona fide} field theory.
The supersymmetric planar model  is exactly equivalent
to the one-dimensional quantum
XXZ Heisenberg chain, and at the same time, to the one-dimensional lattice
gas of q-bosons.
The strong coupling vacua found in the SUSY matrix
model turn out to be nothing but simple ground states of
the XXZ chain which were found more than thirty years ago.
In addition, the XXZ chain appears to have the hidden
supersymmetry which results in many degeneracies among various
energy eigenstates. An unusual feature of these supersymmetry transformations
is that they connect states propagating on different lattices.

Second equivalence is with a gas of q-bosons, with the infinite deformation
parameter. The latter mapping holds exactly in all fermionic and bosonic
sectors while the former does not work for even F and B.

The chain of
correspondences discovered in \cite{VW3} implies also that the specific,
nonlinear $\infty$-bosonic Hamiltonian is in fact soluble via solubility
of the XXZ chain.
{\em Vice versa}: the same solubility implies that the supersymmetric
matrix model is exactly soluble at the infinite value of the 't Hooft coupling.
It is also quite conceivable that this solubility holds in the
whole strong coupling phase \cite{VW2}.

Summarizing: planar calculus applied directly in the Fock space turned
out be rather promising tool in studying some simple, but
non-trivial, models. It allowed to find phenomena which bear a
tantalizing similarity to ones found in the much more advanced
systems  \cite{MZ}-\cite{Ro}. It remains to be seen if this approach
can be extended to the full, i.e. space extended field theories.

\section*{Acknowledgements}
Most of the results reviewed here have been obtained in collaboration with
Gabriele Veneziano.
I thank him for pointing out the practical
advantages of planar calculus in the Hilbert space, and for numerous,
enlightening and stimulating discussions on specific issues.
This work is partially supported by the grant of Polish Ministry of
Science and Education P03B 024 27 (2004)--(2007).

\vspace*{.5cm} A special session during this School is dedicated to
Andrzej Bia\l as in honour of his 70-th birthday. I would like to
thank him for "being there" and for the incredible passion and drive
for physics he is distributing among all of us who have a chance,
and a privilege, to interact with him.

\end{document}